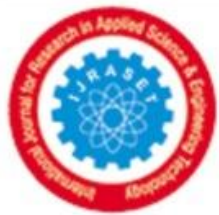

# ijraset

International Journal For Research in
Applied Science and Engineering Technology

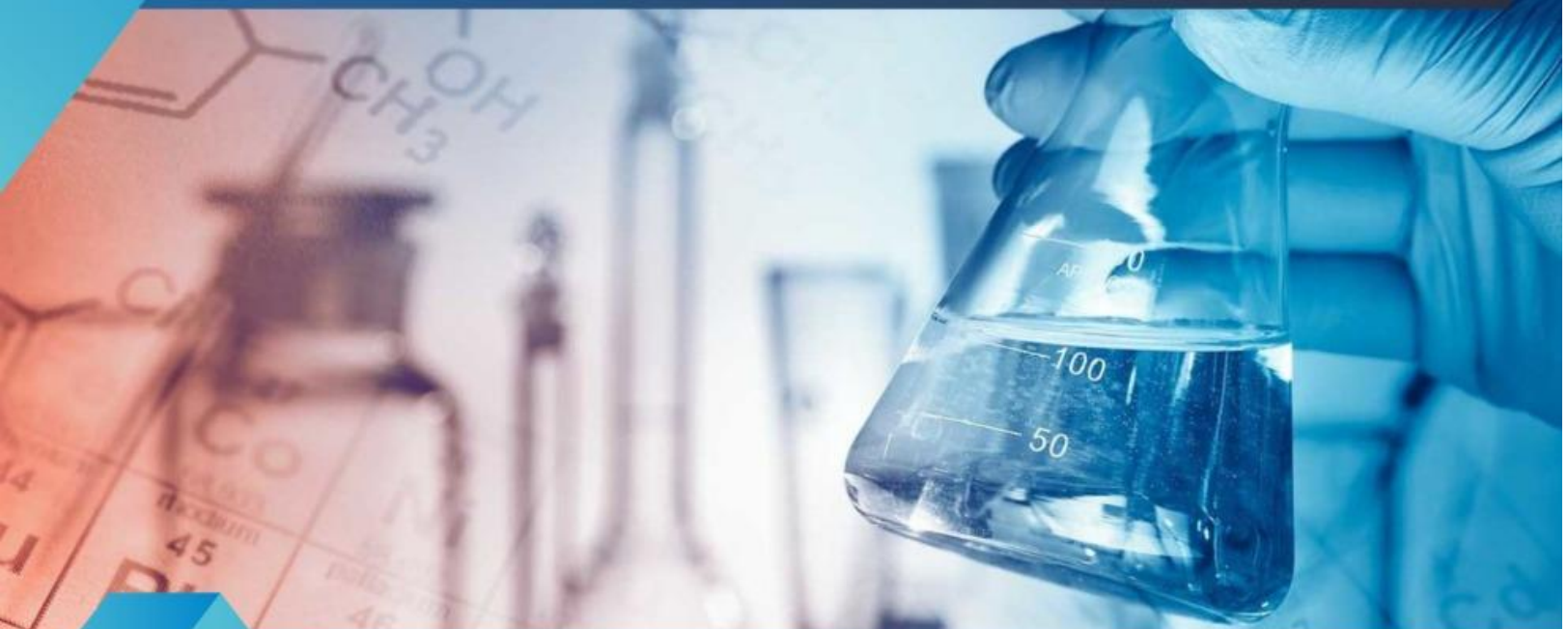

# INTERNATIONAL JOURNAL FOR RESEARCH

IN APPLIED SCIENCE & ENGINEERING TECHNOLOGY

www.ijraset.com

Call: 08813907089    E-mail ID: ijraset@gmail.com

# Design To Convert a Wired PLC into Wireless PLC

Sushil Ghildiyal[1], Kishankumar Bhimani[2], Prof. Manimozhi M[3]
[1, 2, 3] *School of Electrical Engineering, VIT University*

*Abstract*: *This paper implies Bluetooth technology, which is put into effect to alter extant, wired into wireless Programmable Logic Controller (PLC). Here two Bluetooth devices are employed as a transceiver to transmit and receives the input signal to contrive wireless PLC. The main advantage of PLC is to control the output according to the status of input. In Bluetooth technology, the handshaking between the two Bluetooth modules takes place, which is interfaced with a microcontroller board (Arduino board) and then to PLC such that field devices can be controlled without wire.*
*Keywords*: *Bluetooth Module, Arduino Board, Programmable Logic Controller (PLC), Field Bus*

## I. INTRODUCTION

Automation means imparting of human control functions to any technical equipment. Therefore, control panel plays a vital role in an industry; it acts as a brain of any process that commands the desired parameters according to our requirement. Inside the panel, it contains controller either Programmable logic controller (PLC) or other with switchgear etc. Programmable Logic Controller (PLC) is one of the logic controller & solid-state device, which control process or output, based on input and program logic. PLCs read the signals either in digital or analog form from different input devices (sensor, transmitter, keyboard, encoder etc) then as per the program logic it writes it to the output module and then from it, goes to an output device (motors and solenoid valves) to perform the desired function.

In industry, field network plays a major role to run any process. At the field level, I/O devices (Sensors, Motors and Valves etc.) are connected to each other with real-time control networks i.e. Fieldbus. The Fieldbus is able to transmit information related to maintenance and real-time messages, alarms.

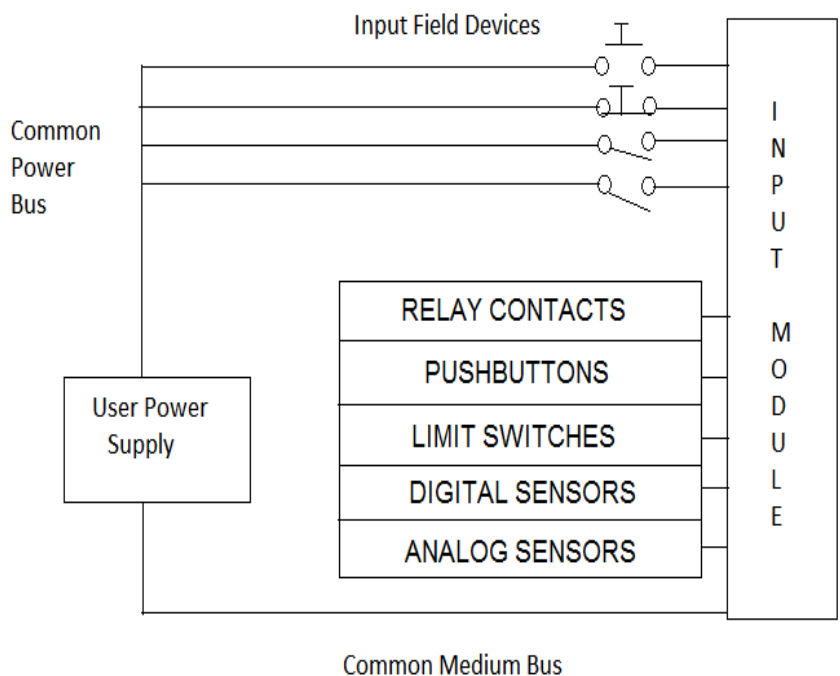

Figure 1 - Input Module

Figure 1 represents the input module and its connection with the input devices. This module receives the signal from input devices and sends it to the controller; controller reads the signal and responds according to the program written.
Figure 2 represents the output module and its connection with output devices and these get actuated as per program.

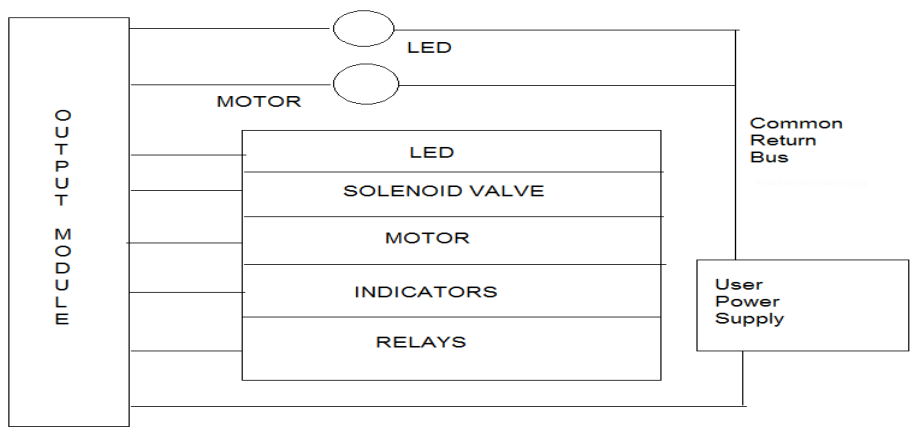

Figure 2 – OUTPUT MODULE

Figure 3 shows the control network of any process. In this network, the field device is connected to PLC with wire and transmitting as well as receiving the signals through Fieldbus.

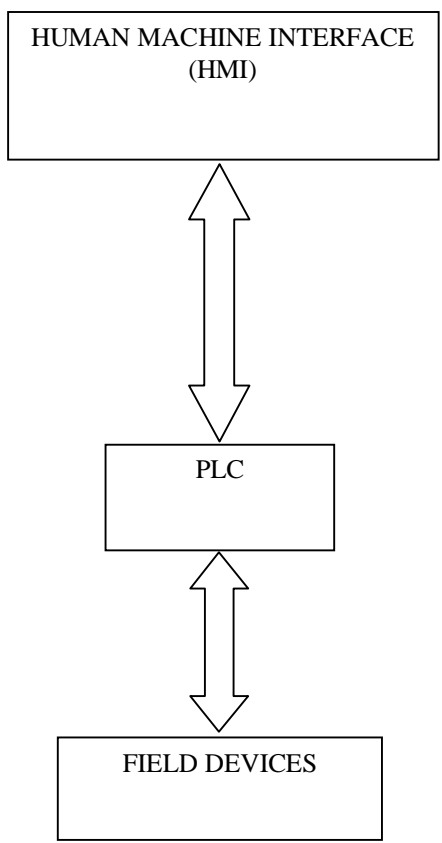

Figure 3 - Control Network

With up gradation in wireless technology, the field devices can be controlled without wire. As in industry through this technology, we can reduce human efforts by reducing the cabling, capital, have less nuisance with connectors. This technology can be used where wiring and troubleshooting is difficult as it saves time and can simultaneously provide a backup communication link in case of normal network failure. As ZigBee is one of the WPAN but it has low data transmission rate. Due to this specification, the scan time for the whole process increases. Hence it is difficult to apply in those areas or processes where with minimum interval pulses of encoder is used to perform a specific task whereas Bluetooth technology is proposed as it is easy to use, of low cost, and of a high transmission rate.

## II. WIRELESS AND WIRED NETWORK

Figure 4 gives a detailed idea about wired and wireless networks. In the table, it focuses on Data transmission rate (kbps), segment length (m) and nodes. Here we can see that for short distance communication Bluetooth technology gives us the good data transmission rate.

A. *In the wireless network, there exist two topologies*
1) *Star topology:* As the name implies the start format for an IEEE 802.15.4 network topology has one central node called the PAN coordinator with which all other nodes communicate.
2) *Peer to peer network topology:* In this type of system topology, there is still what is named a Pan facilitator, yet correspondences may likewise happen between various hubs and not really by means of the organizer. In this paper, we use one of the wireless technology i.e. Bluetooth (IEEE 802.15.1) with PLC to contrivance wireless system with field devices.

TABLE I

|  | Name | Length (m) | Data Transmission rate (Kbps) | Nodes |
|---|---|---|---|---|
| Wire | Profibus | 1200 | 93.75 | 32 |
|  |  | 600 | 182.5 |  |
|  |  | 200 | 500 |  |
|  | Device net | 500 | 125 | 64 |
|  |  | 250 | 250 |  |
|  |  | 100 | 500 |  |
| Wireless | ZigBee | 100 | 250 | 260 |
|  | Bluetooth | 10 | 1000 | 10 |
|  | Wi-Fi | 50 | 5500 | 40 |

## III. DISCUSSION ON BLUETOOTH AND PLC

A. *Bluetooth*

Bluetooth is among one of the wireless technology widely used in day-to-day life because of its low cost, low power. It is easy to use technology which uses frequency radio waves in the (Industrial Scientific and Medical) ISM band (2.4 to 2.485 GHz) to connect, transfer, and share information between various devices. It became very favourable device because of wider application in industry, home appliances, and computer technology. A Bluetooth device/module embraces Bluetooth receiver and transmitter software, which fulfil user-friendly to any person.

When two Bluetooth devices wish to pass on, they use Trans receive signals and hence data pairing. In order to transfer data wirelessly, Bluetooth and Wi-Fi are used. However, both technologies differ in their functioning. Bluetooth holds remote ad-hoc network, which is a disentranced network.

B. *PLC*

A Central Processing Unit (CPU) in technical arena is like the brain of PLC as it performs most of the calculations. As central CPU alludes to the processor, it is very complicated electronic circuitry, which performs the task of executing stored program instructions. CPU has two units: 1) The Control Unit (CU) 2) The Arithmetic Logic Unit (ALU). The control unit does not itself execute the instructions but guides the other parts of the systems in accomplishing this, whereas ALU is used for executing the logical and arithmetic instructions  There is some step performed by CPU as fetch, decode, execute and store.

Figure 5 shows the architecture of PLC as it contains CPU with program memory, data memory, and input/output module and isolation barrier.

Memory gives changeless capacity to the working framework for information utilized by the CPU. The framework's Read Only Memory (ROM) stores information for all time for the Operating System. Arbitrary Access Memory (RAM) stores status data for I/O devices. PLCs either need a device a PC or reassure to download the program into the CPU. There are three types of PLCs i.e. fixed, modular and rack type. We can choose any of these according to our applications.

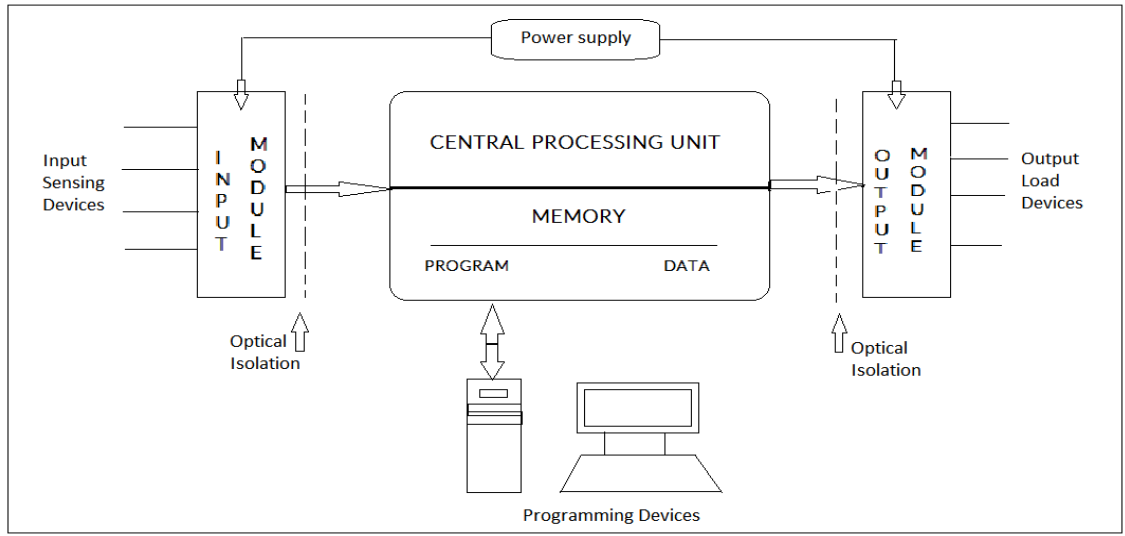

Figure 5 – PLC System

*C. Design And Implementation*

The Figure 6 represents that how the wireless communication establishes between field devices and PLC. Both Bluetooth devices are transceiver that can transmit as well as receive the signal.

The proposed system consists of PLC, Bluetooth module, and Arduino board. The single HC05 acts as a transceiver; therefore, there exists a wireless communication between two HC05 Bluetooth modules, the modules are configured as master and slave. One HC05 module is connected to PLC and other to the field side; therefore, the input signal from sensors, switches or other input devices are first send to arduino, as soon as arduino receives the signal therefore as per code written it transmits the signal to HC05 Bluetooth module connected with it, which then transmits the signal to the other Bluetooth module which is connected to the PLC side. PLC receives the information from Bluetooth module and actuates the output as per program. The signal transmission between two input device and PLC is performed without wire.

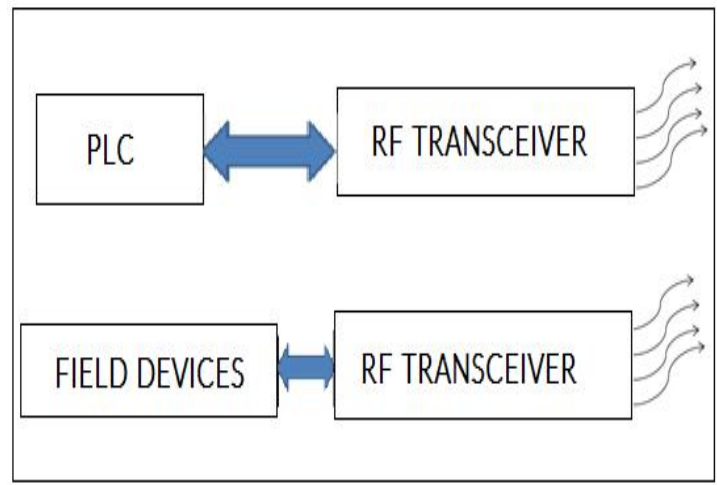

Figure 6 - Transceiving method

*D. Components And Tools*

The Components required for this model are
1) PLC: 14SS DELTA with 8 Inputs and 6 Outputs channels. As this PLC is cheap, easy to use, easy to communicate, and widely used in industry. The operating voltage for this PLC is 24VDC and inputs to the I/O channel are 24VDC.

*2)* Bluetooth and Arduino: HC05 Bluetooth shown in Figure 7 with Arduino Uno board. Since Arduino Uno is an open source platform therefore easy to use.
*3)* Relay: It is an electromechanical device with coil voltage is 5VDC; use to transmit 24VDC to PLC input module as relay holds.

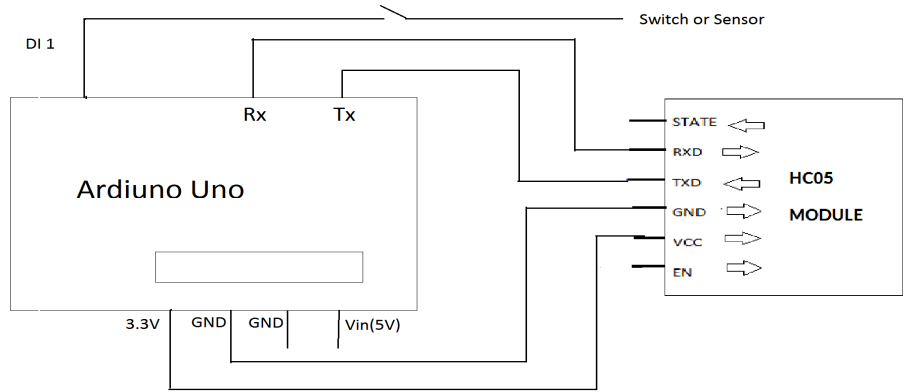

Figure 7 - Arduino Uno & HC05 Bluetooth module Interfacing

The software used to program PLC is WPL Software 2.39 is Windows-based and programming language used is Ladder Block Diagram (LBD). Figure 8 represents the Arduino Programming software i.e Arduino IDE.

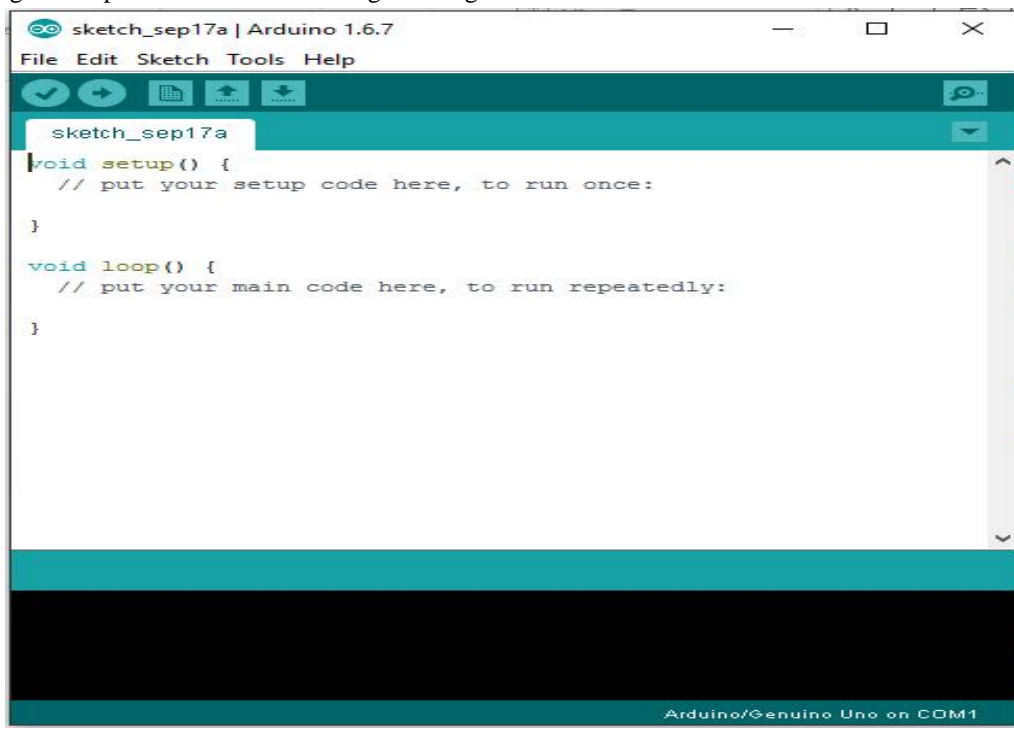

Figure 8 - Arduino IDE

### IV. BUILDING THE SYSTEM

Figure 9 given below shows the field side circuitry as here we have taken switch as an input device which is connected to the controller (Arduino) and Arduino is interfaced with HC05 (slave) Bluetooth module with six pins as EN,TXD, RXD, Vcc, GND and STATE.
In order to execute HC05 modules as master or slave, we have to set it to AT-Command modes with ROLE 1 for master and 0 for the slave. Both master and slave module should have same baud rate, and the address of slave module have to Bind up with master module through AT+BIND='address of the slave module' command.

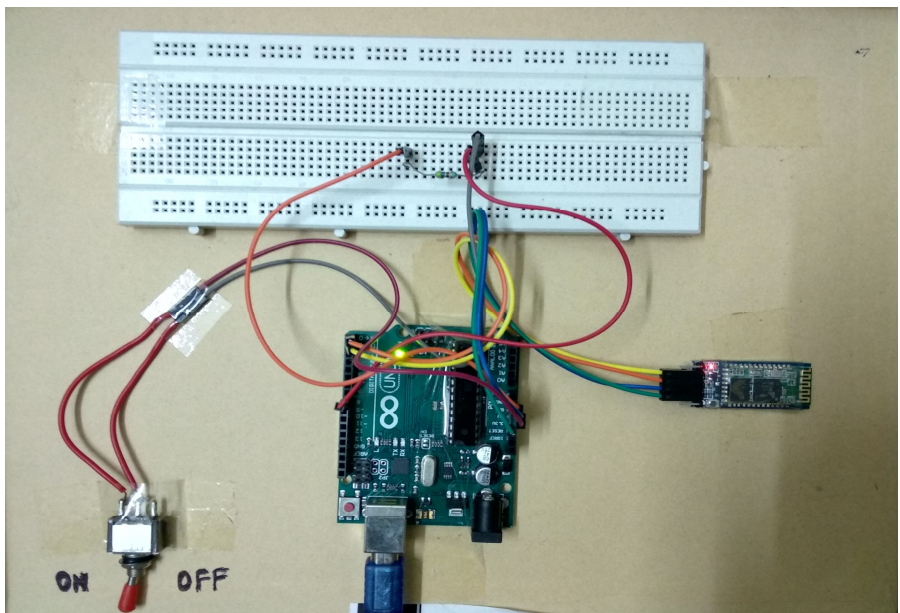

Figure 9 – Field Side

This implies that as we switch ON, the handshaking between master/slave modules takes place and a wireless communication established between two modules. The signal through slave module transmit to master module wirelessly and according to the Arduino code it makes the digital input high which then holds the relay, and due to this holding of a relay, the signal gets transmitted to the input channel of the PLC which results in making the output high according to our program.

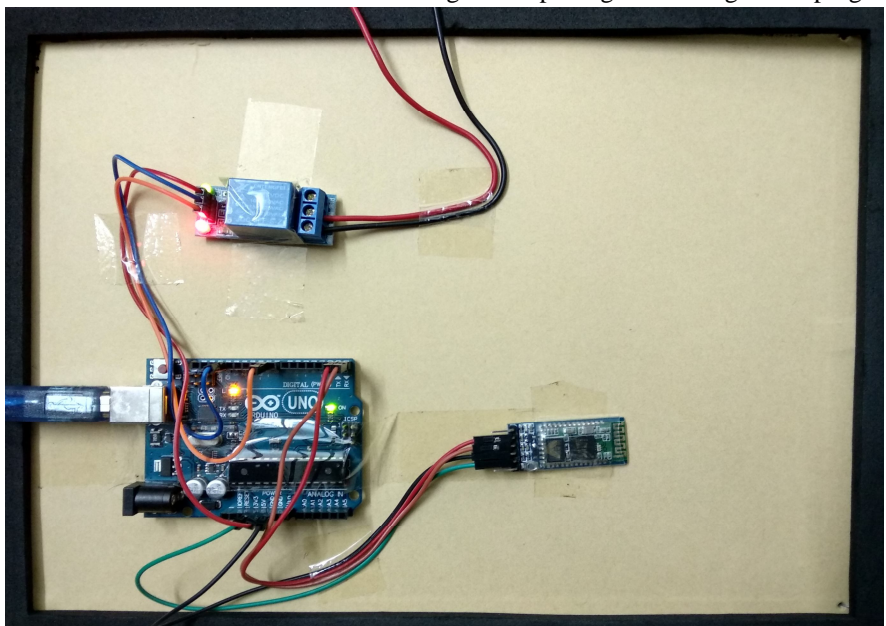

Figure 10 – HC05 Master Module

Figure 10 shows HC05 master module connection with controller (Arduino Uno) and relay acts for switching of 24VDC to input side of the PLC.

Figure 11 shows Delta PLC and power supply used. This figure shows the output state when input is high. Here it is programmed as whenever the switch is ON the output has to be high and vice versa therefore, this figure shows that the output Y1 is high when switch is ON.

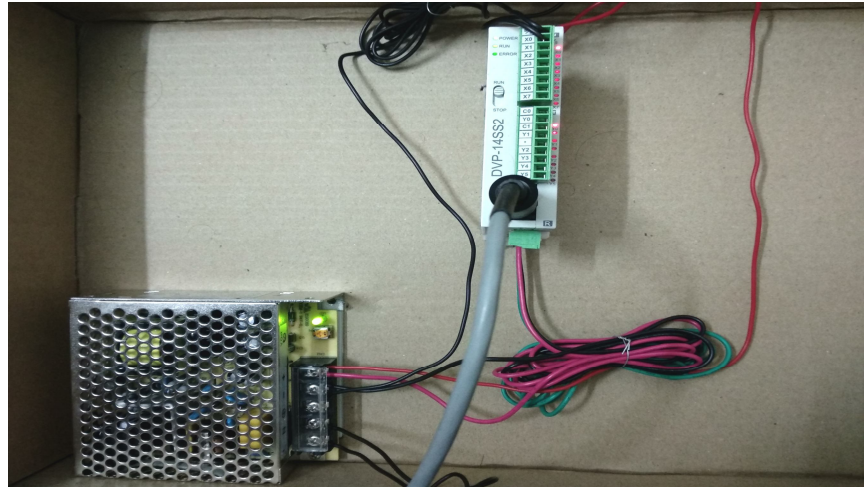

Figure 11 – Delta PLC

## V. CONCLUSION

In this paper, based on the HC05 Bluetooth module a wireless fieldbus for PLC is designed which can be controlled without wire.

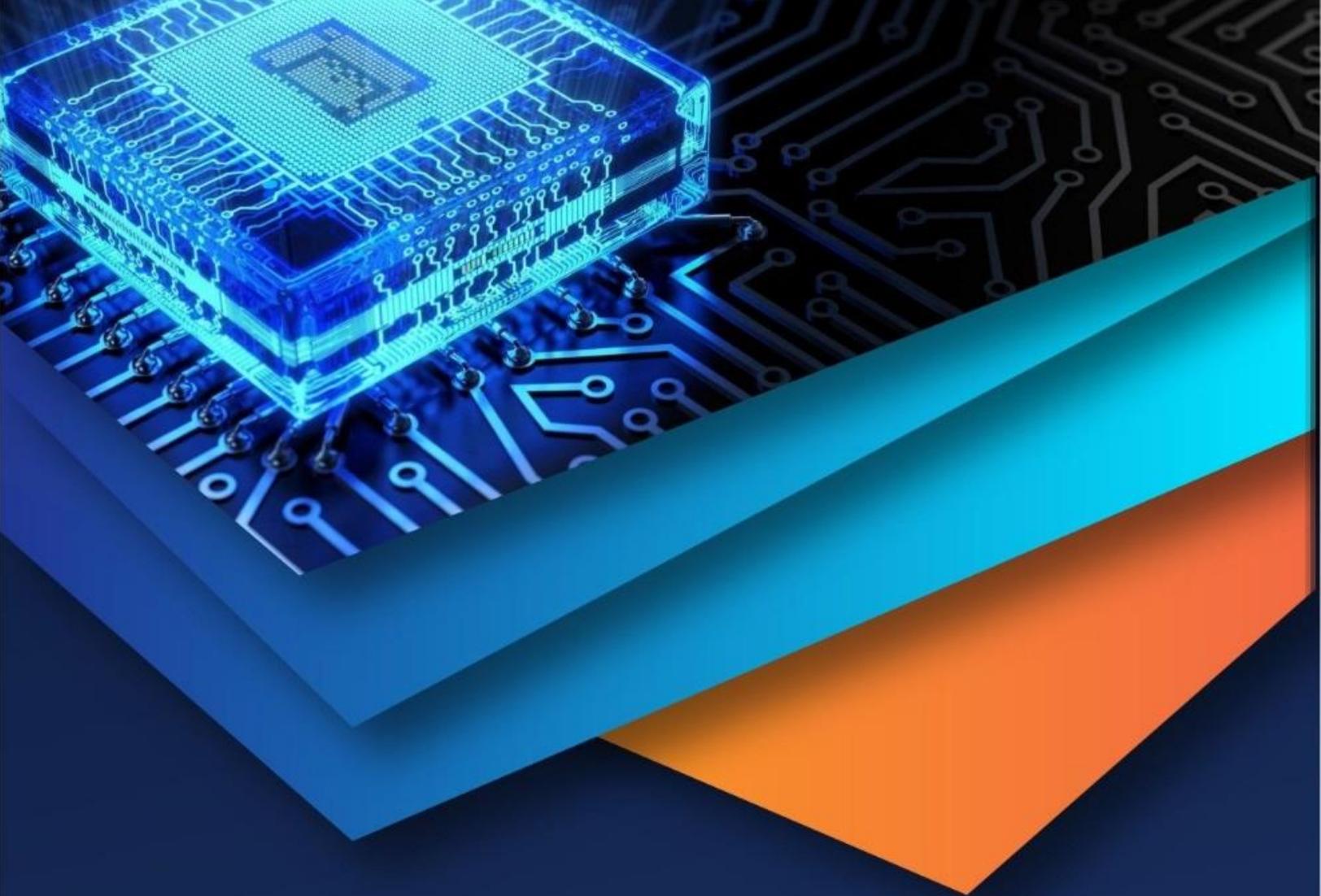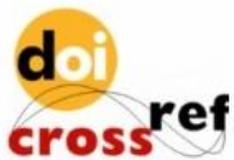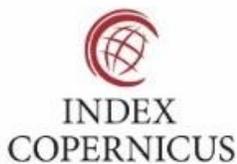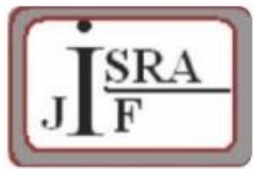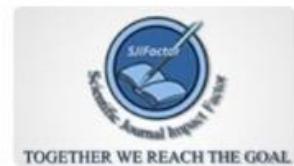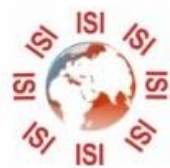

# INTERNATIONAL JOURNAL FOR RESEARCH

IN APPLIED SCIENCE & ENGINEERING TECHNOLOGY

Call : 08813907089 (24*7 Support on Whatsapp)